\documentclass{aa}

\usepackage[utf8]{inputenc}
\usepackage{graphicx}
\usepackage{natbib}
\usepackage{array}
\usepackage{makecell}

\begin{document}

\title{A new method for estimating global coronal wave properties \\ from their interaction with solar coronal holes}

\titlerunning{Global coronal waves properties from their interaction with CHs}
\authorrunning{Piantschitsch, I., Terradas, J., Temmer, M.}

\author{Piantschitsch, I.$^{1,2}$, Terradas, J.$^1$, Temmer, M.$^2$}

\institute{$^1$Departament de F\'\i sica, Universitat de les Illes Balears (UIB),
E-07122, Spain \\   Institute of Applied Computing \& Community Code (IAC$^3$),
UIB, Spain\\
$^2$Institute of Physics, University of Graz, Universit\"atsplatz 5, A-8010 Graz, Austria
\\
\email{isabell.piantschitsch@uib.es}
}

\date{}

\abstract{Global coronal waves (CWs) and their interaction with coronal holes (CHs) result, among other effects, in the formation of reflected and transmitted waves. Observations of such events provide us with measurements of different CW parameters, such as phase speed and intensity amplitudes. However, several of these parameters are provided with only intermediate observational quality,
%such as measurements of kinematics of the reflected waves, 
other parameters, such as the phase speed of transmitted waves, can hardly be observed in general. We present a new method to estimate crucial CW parameters, such as density and phase speed of reflected as well as transmitted waves, Mach numbers and density values of the CH's interior, by using analytical expressions in combination with basic and most accessible observational measurements. The transmission and reflection coefficients are derived from linear theory and subsequently used to calculate estimations for phase speeds of incoming, reflected and transmitted waves. The obtained analytical expressions are validated by performing numerical simulations of CWs interacting with CHs. This new method enables to determine in a fast and straightforward way reliable CW and CH parameters from basic observational measurements which provides a powerful tool to better understand the observed interaction effects between CWs and CHs.}

\keywords{Magnetohydrodynamics (MHD) --- waves --- Sun: magnetic fields}

\maketitle

\section{Introduction}

Coronal waves (CWs) are large scale propagating disturbances in the corona and considered as fast mode magnetohydrodynamic (MHD) waves \citep[see e.g.,][]{vrsnaklulic2000}. Evidence for their wave characteristics is given from observations of secondary waves when interacting with coronal holes (CHs) representing regions of sudden changes in density, Alfv\'en and magnetosonic speed. Secondary waves are caused by reflection and refraction at the boundary of a CH \citep[e.g.,][]{kienreichetal2013, Long2008, gopal2009} or transmission through a CH \citep{olmedoetal2012,liu19}. Case studies of chromospheric Moreton waves also show a partial penetration into a CH \citep[e.g.,][]{Veronig2006}. Numerical simulations confirm the wave interpretation in accordance with the observations by finding effects such as deflection, reflection and refraction when the wave interacts with a structure like a CH \citep{afanasyev2018,Piantschitsch2018a, Piantschitsch2018b}.

Typical wave parameters of primary and secondary wave fronts that can be measured using observations are phase speed, intensity amplitude and width \citep[e.g.,][]{Muhr2011,Kienreich2011}. However, secondary and especially transmitted waves are rather weak in their signal, hence, quality and accuracy of measurements are rather low which might lead to a misinterpretation of the results. With that also other coronal parameters giving information e.g., about the CH itself, are difficult to derive. In particular, information about dynamics and density distribution inside of a CH is mostly unavailable due to the CH's low density compared to the surrounding area.
%\textcolor{blue}{ In particular, information about dynamics and density distribution inside of a CH is mostly unavailable due to the CH's low density compared to the surrounding area.} 
%This lack of crucial information about CW parameters and CHs \textcolor{blue} {leads to difficulties in the description and interpretation of many interaction effects, such as coronal dimming, the evolution of rarefaction regions or kinematics of transmitted waves.} 
Numerical simulations are capable of providing additional information about CW parameters and the interaction effects between CWs and CHs but are still limited considering their necessary idealization and dependence on initial conditions. %which anyway are usually taken from observational measurements.

The aim of this Letter is therefore, to provide a method to determine estimations for important CW parameters, such as density amplitudes of reflected and transmitted waves, by using simple analytical terms in combination with basic observational measurements. These estimations will be obtained by first, deriving analytically reflection and transmission coefficients from linear theory, and second, complementing these terms with simple expressions of nonlinear MHD waves. We are going to validate the theoretical expressions by performing numerical simulations of CW propagation and its interaction with CHs. For the comparison of the theoretical results with observations we chose two different events of CW-CH-interaction, which differ in their phase speed of the secondary waves. The first case represents a purely acoustic case where the phase speed is close to the typical sound speed \citep{kienreichetal2013}.The other case will be used to validate the theoretical expressions determined for the purely magnetic case, due to the fact that the observed phase speed of the incoming wave is close to $700$ km $\rm s^{-1}$ \citep{olmedoetal2012}. Overall, we will show that this newly developed method is a useful and fast tool to provide important information about CW parameters, dynamics inside a CH as well as interaction effects between CWs and CHs.

%{\bf In Section 2 we introduce the basic model and the initial setup we are %going to use for the parameter estimations and the numerical experiments. %Moreover, we present the derivations of estimations for the coronal wave %parameters in the linear and nonlinear case and compare them to the results of %the numerical simulations. In Section 3 we apply the theoretical and numerical %results to observations of a purely acoustic case \citet{kienreichetal2013} as %well as for a purely magnetic case \citet{olmedoetal2012}. We conclude in %Section 4.}

%{\bf \color{red}
%Not sure about this paragraph. I think it is not necessary in a Letter and it is %a sort of summary of the previous paragraph...}

\section{Theoretical results}\label{theory}

In this section we first analytically derive reflection and transmission coefficients from linear theory and validate these terms by performing numerical simulations of CW-CH-interaction. Second, we use these coefficients to obtain analytical expressions for the phase speeds of incoming, reflected and transmitted waves.

\subsection{Linear case}\label{lincase}

We start with a simple equilibrium based on uniform density, $\rho_0$, gas pressure, $p_0$, and a magnetic field, $B_0$, pointing in the $z-$direction. We focus on perturbations propagating in the $x-$direction perpendicularly to the magnetic field, representing CWs. Since 
the equilibrium is homogeneous, sound ($c_{\rm s0}=\sqrt{\gamma p_0/\rho_0}$) and Alfv\'en speeds ($v_{\rm A0}=B_0/\sqrt{\mu_0 \rho_0}$) are constant, and fluctuations in the system propagate as plane waves. For the velocity we have
\begin{equation}\label{simplewave}
    v=v_0\,e^{i\left(\omega t \pm k_x x\right)}.
\end{equation}
By using this expression in the standard linearised MHD equations and performing the temporal and spatial derivatives, a dispersion relation is readily obtained
\begin{equation}\label{disper}
    \omega=k_x\, c_{f0}=k_x \sqrt{c_{s0}^2+v_{A0}^2},
\end{equation}
which corresponds to fast MHD waves propagating purely perpendicular to the magnetic field at the fast speed ($c_{f0}$). The density changes due to these waves in the linear regime are given by the simple expression
\begin{equation}\label{denslinva}
    \rho=\rho_0\left(1\pm \frac{v}{c_{f0}}\right),
\end{equation}
for right (+) and left (-) propagating waves. 

In the low-$\beta$ situation, we neglect the sound speed since it is negligibly small compared to the Alfv\'en speed and the dispersion relation reduces to $\omega=k_x\, v_{A0}$. In the high-$\beta$ limit the magnetic field is very weak and we obtain the dispersion relation of purely acoustic waves, $\omega=k_x\, c_{s0}$. 

\begin{figure}[t!]
\includegraphics[width=\linewidth]{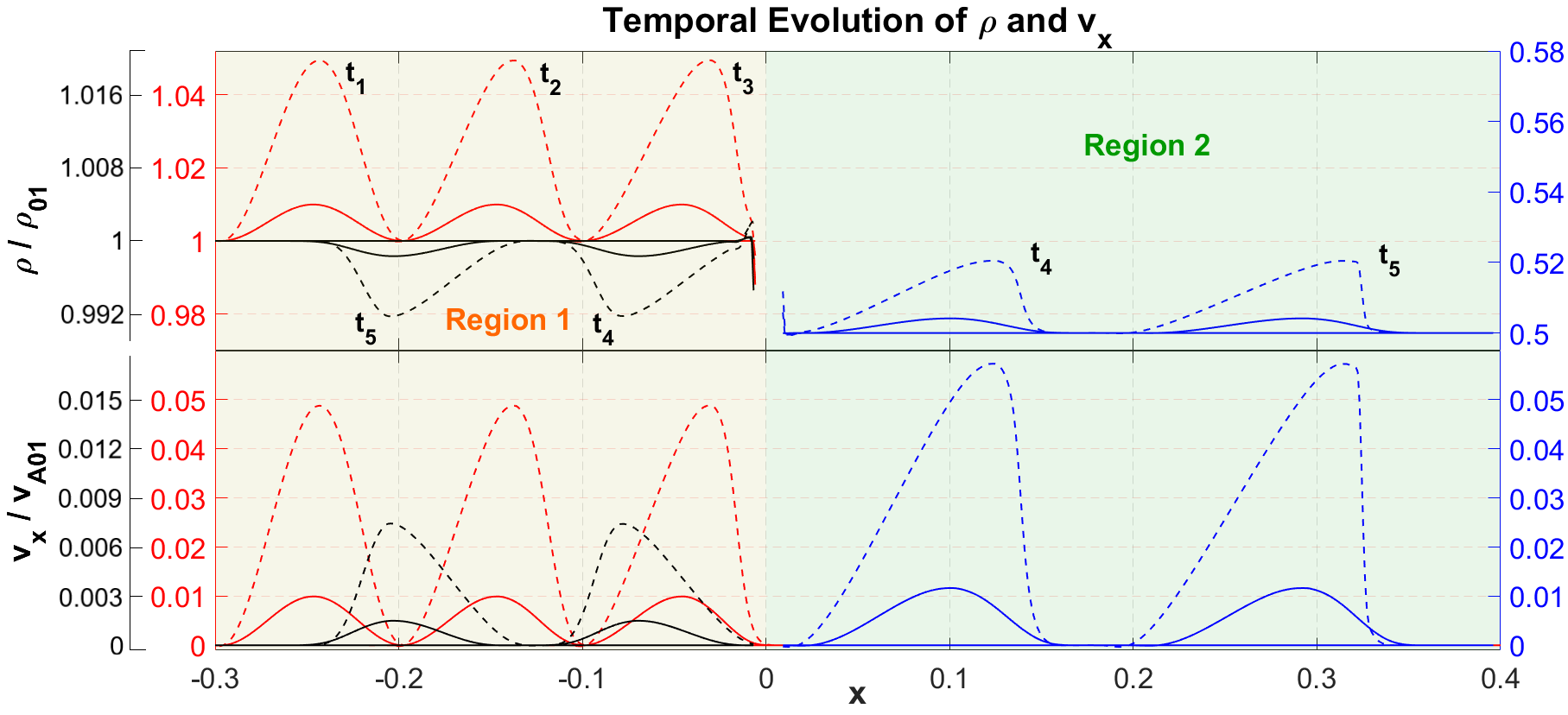}
\caption{\small Density (top panel) and velocity (bottom panel) at five different times during the evolution of a linear (solid line) and weakly nonlinear (dashed line) perturbation representing an idealised CW. The incoming wave (red) steepens into a shock in the nonlinear regime. The reflected wave at the interface between Region 1 and Region 2, which represents the CH boundary (located at $x=0$) is a rarefaction wave (black) while the transmitted (blue) is a shock wave.}\label{retefig1}
\end{figure}

We now extend the situation for a homogeneous medium to the interface problem which is based on two different homogeneous media connected through a discontinuity at $x=0$ (with densities $\rho_{01}$ in region 1 and $\rho_{02}$ in region 2, see top panel in Fig.~\ref{retefig1}). This is an idealized representation of a CH (corresponding to region 2 with $\rho_{02}<\rho_{01}$) but allows to consider the basic properties of the reflection/transmission of a fast MHD wave at a density step.

We assume again a plane wave propagating in region 1, which interacts with the interface and generates a reflected wave at the CH boundary ($x=0$). The incoming and reflected waves at the interface are of the form
\begin{equation}\label{incoreflect}
    v_1=v_0\,e^{i\left(\omega t- k_{x1} x\right)}+v_R\,e^{i\left(\omega t+ k_{x1} x\right)},
\end{equation}
In region 2 there is a transmitted wave traveling through the CH
\begin{equation}\label{transmit}
    v_2=v_T\, e^{i\left(\omega t- k_{x2} x\right)}.
\end{equation}
In this problem the frequency $\omega$ is constant but the wave number changes according to the dispersion relation in the corresponding medium. For this reason we have now two wavenumbers, $k_{x1}$ and $k_{x2}$. Again, if the sound speed is negligible compared to the Alfv\'en speed we obtain $k_{x1}=\omega/v_{A01}$ and $k_{x2}=\omega/v_{A02}$. The amplitudes of the velocities in the two regions are not independent from each other since the variables have to satisfy certain conditions at the interface \citep[see for example,][]{walker2004}. In particular, the velocity and the total pressure have to be continuous at $x=0$ (the location of the interface). When these conditions are fulfilled it is straight forward to obtain the following amplitudes
\begin{equation}\label{sr}
    v_R=\frac{\sqrt{\rho_{01}}-\sqrt{\rho_{02}}}{\sqrt{\rho_{01}}+\sqrt{\rho_{02}}} v_0=\frac{1-\xi}{1+\xi} v_0,
\end{equation}
which correspond to the reflection coefficient as a function of the incoming wave amplitude, $v_0$, and 
\begin{equation}\label{st}
    v_T=\frac{2 \sqrt{\rho_{01}}}{\sqrt{\rho_{01}}+\sqrt{\rho_{02}}} v_0
    =\frac{2}{1+\xi} v_0,
\end{equation}
for the transmission coefficient, where the density contrast is defined as $\xi=\sqrt{\rho_{02}/\rho_{01}}$ ($0<\xi<1$ for CHs). Note that the coefficients satisfy the equation $v_0+v_R=v_T$ and always have the sign of $v_0$. 

 Let us assume that the velocity amplitude of the incoming wave, $v_0$, is positive. This means that, according to Eq.~(\ref{denslinva}), the incoming wave has a density enhancement associated to it ($\rho>\rho_0$). On the contrary, the reflected wave, with a negative sign in Eq.~(\ref{denslinva}) but being $v_R>0$ (because $v_0>0$, see Eq.~(\ref{sr})) corresponds to a density dimming ($\rho<\rho_0$). This is in agreement with the reported reflections of CWs at CHs by most of the observations \citep[e.g.,][]{kienreichetal2013, gopal2009}.

Finally, it turns out that for pure sound waves the reflection and transmission coefficients are exactly the same as for the purely magnetic case, therefore Eqs.~(\ref{sr}) and (\ref{st}) are used in the acoustic case as well. In this last situation the linear density fluctuations are given by Eq.~(\ref{denslinva}) but with $c_{f0}$ replaced by $c_{s0}$.

%Besides the changes in the amplitude of the wave the wavenumber also varies due to the different properties of %the medium. This means that if instead of a monochromatic wave with a single wavenumber we have  a pulse which %can be always decomposed as a superposition of discrete wavenumbers, the width of the transmitted pulse will %also change. For a constant magnetic field we have that 
%\begin{equation}\label{kcoef}
%    \frac{k_2}{k_1}=\sqrt{\frac{\rho_{02}}{\rho_{01}}}.
%\end{equation}
%Hence, if the density of the second region, $\rho_{02}$ is smaller than $\rho_{01}$ as it happens for a %coronal hole, then according to Eq.~(\ref{kcoef}) $k_2$ is smaller than $k_1$ meaning that the wavelength is %larger in the second medium ($\lambda=2\pi/k$). The pulse is increasing its width as it crosses the interface.

\subsection{Numerical experiments in the linear regime}\label{numexp0}

%{\bf Isabell:} {\em In this subsection I would like you to include a plot of the %theoretical reflection/transmission coefficients together with the amplitudes %inferred from the simulations. You could represent the curves for different %density values as a function of the initial velocity amplitude. For small %amplitudes the agreement is very good but there are discrepancies, not so big, %around $12\%$, for larger amplitudes, we will make reference to this part in the %following sections.
%}

%\color{red}
%I'm not sure about the structure and the position in the letter of this section. %The numerical and theoretical part is a little bit messy concerning the %structure.

%\color{blue}
%The model we use to derive the estimations for the coronal wave parameters and %to perform numerical simulations of coronal wave propagation includes a vertical %and constant magnetic field, ${\bf B}=B\, {\bf \hat{e}}_z$ pointing in the %$z-$direction as well as a separation of the density domain into two different %areas ($\rho_{1}$ and $\rho_{2}$). We excite an initial perturbation for the %parameters $\rho$, $v_{x}$ and $B_{z}$, which look in detail as follows:
%\vspace{10px}

%{\bf Equations for initial parameters} \color{red}{(need to be discussed)}
%\vspace{10px}

\begin{figure}[t!]
\includegraphics[width=\linewidth]{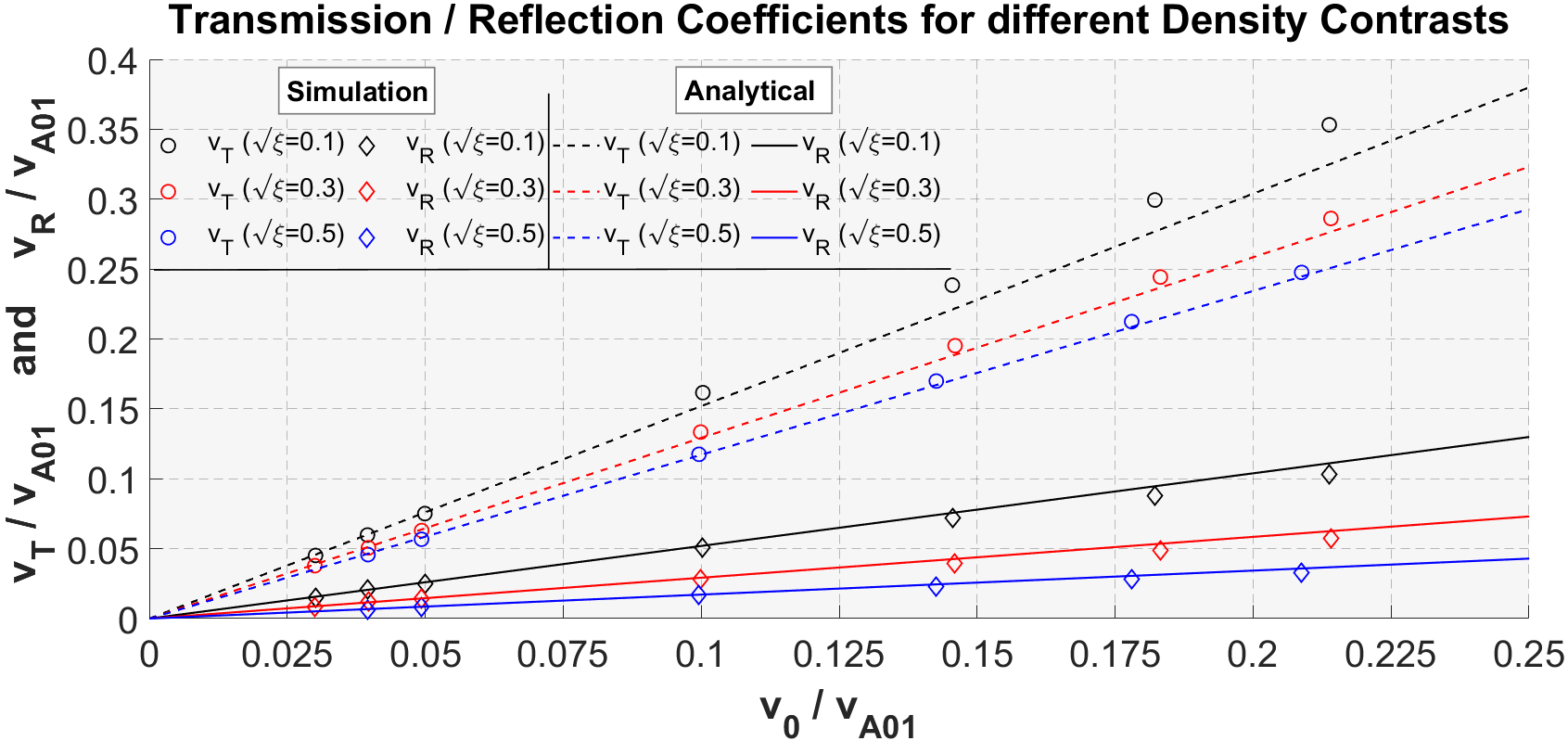}
\caption{\small Transmitted (circles) and reflected velocity amplitudes (diamonds) as a function of the incident velocity amplitude ($v_0$) inferred from the linear and nonlinear MHD simulations for different density contrasts. The dashed and solid lines correspond to the predicted analytical results using the transmission and  reflection coefficients given by Eq.~(\ref{st}) and Eq.~(\ref{sr}).}\label{retefig2}
\end{figure}

Now the MHD wave propagation and its interaction with a region of lower density like a CH is solved numerically by using the standard ideal MHD equations \citep[see][for details]{Piantschitsch2017,Piantschitsch2018a,Piantschitsch2018b}. An initial Gaussian linear perturbation is introduced in the system and the evolution of this fluctuation is followed in time, see Fig.~\ref{retefig1}. This Gaussian pulse can be interpreted as a superposition/combination of different plane harmonic waves, therefore the linear analysis performed in Sect.~\ref{lincase} can be applied to these simulations. The incident wave, corresponding to a density enhancement, see solid red lines in the top panel, travels toward the right and eventually interacts with the density discontinuity at $x=0$. A reflected density dimming, with a low amplitude, is then traveling to the left (solid black line) while a density enhancement is moving at a faster speed towards the right inside the CH (solid blue line). Similar results are found for the velocities, see solid lines in the lower panel of Fig.~\ref{retefig1}. These are the expected results from linear theory. From the simulations we are able to derive the reflected and transmitted velocity amplitudes, see circles in Fig.~\ref{retefig2}, and compare them to the analytical expressions for the reflection and transmission coefficents which we derived in Sect. \ref{lincase} The values obtained from the simulations show good agreement with the theoretical calculations in the linear regime ($v_0\ll v_{A01}$).

%The numerical scheme of the code we use is based on the so-called Total %Variation Diminishing Lax-Friedrichs (TVDLF) method, which is a fully explicit %scheme and was first described by \cite{toth1996}. \color{red} (How many details %of the scheme/setup do you want? Limiter, boundary conditions, gridsize %??$\rightarrow{NONE!!}$)

%\color{red}
%Instead of a plot of the initial setup I would suggest to include here (or at %another position of the paper) a small plot that shows the morphology of the %wave propagation for $\rho$ and $v_{x}$ (i.e. $\rho$ and $v_{x}$ versus position %for different time steps .... the plot you anyway wanted to have). Without such %a morphology plot it's maybe not super clear for the reader what we are exactly %talking about. I saw you would like to have this plot in section of Numerical %experiments. I'm just not sure if we should leave the model description and %numerical setup without any further plot or information. {\color{blue} \bf %Agreed!}

\subsection{Nonlinear case}\label{nonlinsec}

There are evidences that CWs which are propagating in the corona and interacting with CHs are of nonlinear nature \citep{vrsnaklulic2000,Warmuth2004}. %The main goal here is to understand well the behaviour of a nonlinear fast mode wave, especially how the phase speeds of the propagating waves are modified by the presence of nonlinearities, an effect that is reported in the observations.
For pure sound waves ($v_{A0}=0$) the nonlinear results are well known and can be found, for example, in \citet{mihalas1984,landaulif1987}. The nonlinear wave with a velocity amplitude $v$ modifies the local sound speed, $c_{s}$, which is now different from the unperturbed reference sound velocity, $c_{s0}$. For the phase speed of the nonlinear wave we obtain $v_{\rm p}(v)=v\pm c_{s}$ 
and it can be shown that it reduces to the simple expression
\begin{equation}\label{vphasecs}
    v_{\rm p}(v)=\frac{\gamma+1}{2}v\pm c_{s0},
\end{equation}
where we have again the distinction between right and left propagating waves. The fact that the phase velocity of the wave depends on $v$ leads to the steepening of the wave. Now density variations in the nonlinear wave are related to velocity through the following equation under adiabatic conditions
\begin{equation}\label{densnonlincs}
    \rho=\rho_0\left(1\pm \frac{\gamma-1}{2}\frac{v}{c_{s0}}\right)^{2/(\gamma-1)}.
\end{equation}
 %Note that we do not know how $v$ depends on position and time in the previous expressions, and therefore the details of the steepening of the wave are unknown, but this is not relevant for the method developed here.

For the purely magnetic fast wave ($c_{s0}=0$) it is straight forward to derive equivalent equations by exchanging the adiabatic condition with the magnetic induction equation. It can be shown that in this case and for perpendicular propagation \citep[see some details in][]{mann1995} the phase speed is $v_{\rm p}(v)=v\pm v_{A}$. Again it is not difficult to show that this expression reduces to
\begin{equation}\label{vphase}
    v_{\rm p}(v)=\frac{3}{2}v\pm v_{A0}.
\end{equation}
This equation is rather simple since we have eliminated the Alfv\'en speed modified by the presence of the wave ($v_A$), and it involves only the unperturbed Alfv\'en speed ($v_{A0}$) and the velocity amplitude of the wave ($v$). It can be shown that now density variations in the wave are related to velocity through the following equation
\begin{equation}\label{densnonlin}
    \rho=\rho_0\left(1\pm \frac{1}{2}\frac{v}{v_{A0}}\right)^2.
\end{equation}
\noindent In the limit $v/v_{A0}\ll 1$ this expression leads to the linear result of Eq.~(\ref{denslinva}). As noted by \citet{mann1995} the expressions for the magnetic case are simply obtained by setting $\gamma=2$ and replacing $c_{s0}$ by $v_{A0}$ in the equations for the nonlinear acoustic case.

When acoustic and magnetic effects are combined together the problem is more difficult and it requires a numerical treatment, which is out of the scope of this Letter.

\subsection{Numerical experiments in the nonlinear regime}\label{numexpnonlin}

%{\color{red} \bf Isabell:} {\em Not sure if this subsection is necessary, %probably not.} {\color{red} But we can make reference to Plot 1, to see some %nonlinear effects on the waves.}

Here we extend the results of Sect.~\ref{numexp0} to the nonlinear situation. The amplitude of the initial Gaussian perturbation is increased, meaning that the velocity of the wave just before it interacts with the CH ($v_0$ in our notation) is larger than in the linear case. Now the wave shows some steepening as it is approaching the CH (see Fig.~\ref{retefig1}, red dashed line in region 1 of the lower panel) and also once it is transmitted through the CH (blue dashed line). The velocity amplitude of the wave is larger inside the CH and the width of the pulse has increased. The behaviour of the signal in each region follows the behaviour predicted in Sect.~\ref{nonlinsec}, and the equations for the density as a function of velocity are exact. 
But more important, even beyond the linear regime there is still a good agreement with the reflection/transmission amplitudes based on the linear calculations, see circles in Fig.~\ref{retefig2} for $v_0/v_{A01} > 0.1 $. In this figure we can see that for a density contrast of $\xi=\sqrt{0.5}$ the transmitted amplitude agrees quite well with the predicted linear value (see dashed blue line). The differences become more prominent the smaller the density contrast, however, in the worst case the error is only around 12$\%$. This has important consequences for the method developed in this work, meaning that the simple linear expressions for reflection and transmission coefficients can be used to estimate important CW parameters which are originally nonlinear in nature.

\subsection{Implications}

%We first discuss the implications of the detection of two different velocities in the incoming and reflected wave at the CH boundary. 
The measured phase speed of the incoming front in the observations is denoted by $u_I$ while the measured reflected phase velocity is $u_R$. According to the previous equations for the purely magnetic case we have that (see Eq.~(\ref{vphase})) for the incoming wave
\begin{equation}\label{ui}
    u_I =  v_{A01}+\frac{3}{2}v_0,
\end{equation}
while for the reflected wave (we implicitly assume that it is propagating to the left and the global minus sign is not taken into account)
\begin{equation}\label{ur}
    u_R =  v_{A01}-\frac{3}{2}v_R.
\end{equation}
%According to these expressions we have to expect that $u_R$ is always smaller than $u_I$, because of the minus sign in the second term of Eq.~(\ref{ur}) and also because the reflected velocity amplitude $v_R$ is supposed to be smaller than $v_0$, the amplitude associated to the incoming wave (this is true when we recall the reflection coefficient given by Eq.~(\ref{sr})). Indeed the observations indicate that $u_R<u_I$.

By combining Eq.~(\ref{ui}) and Eq.~(\ref{ur}) we obtain
\begin{equation}\label{urn}
    u_R  =  u_I -\frac{3}{2}\left(v_0+v_R\right)=u_I-\frac{3}{2}v_T.
\end{equation}
Remember that $v_R$ and $v_T$ are simply the reflected and transmitted amplitudes given by Eqs.~(\ref{sr})-(\ref{st}). Although we are dealing with nonlinear waves the linear results about the reflection/transmission problem are still applicable, see Sect.~\ref{numexpnonlin}, this is a key point of the method presented here. We do not need to apply the Rankine-Hugoniot jump conditions.

\section{Application to observations}

In this section we apply the analytical expressions we obtained in Sect.~\ref{theory} to observations and compare the results to two different case studies which differ in the phase speed measurements of the propagating CW.

%{\color{red} For the first event we use the theoretical expressions for %the purely acoustic case and compare them to the results of %\citet{kienreichetal2013} where we find measured coronal wave phase %speed that are close to the local sound speed. In the second event we %make use of the expressions obtained for the purely magnetic case and %compare them to the measurements in \citet{olmedoetal2012} where phase %speeds of around $700$ km $\rm s^{-1}$ are reported.}

%{\color{red} \bf Is this paragraph useful?, it is a sort of repetition %of the information that it is written below for Events 1 and 2..., I %would remove it...}

It is straightforward to use the transmitted amplitude in Eq.~(\ref{urn}) to obtain the velocity amplitude of the incoming wave in terms of the density contrast, $\xi=\sqrt{\rho_{02}/\rho_{01}}$, and the phase velocities of incident and reflected waves
\begin{equation}\label{v0est}
    v_0  =\frac{1}{3}\left(1+\xi\right)
    \left(u_I-u_R\right).
\end{equation}
Therefore, we are able to calculate the velocity amplitude of the front close to the CH boundary, a magnitude that the observations are unlikely to provide. With this information and using Eq.~(\ref{ui}) the local Alfv\'en speed is simply 
\begin{equation}\label{vA0est}
    v_{A01}=\frac{1}{2}\left(u_I(1-\xi)+u_R(1+\xi)\right).
\end{equation}

\noindent Repeating the same derivation but for the purely acoustic case we now find that
\begin{equation}\label{v0estacoust}
    v_0  =\frac{1}{\gamma+1}\left(1+\xi\right)
    \left(u_I-u_R\right),
\end{equation}
while the background sound speed is given by
\begin{equation}\label{cs0est}
    c_{s01}=\frac{1}{2}\left(u_I(1-\xi)+u_R(1+\xi)\right),
\end{equation}
which is completely equivalent to Eq.~(\ref{vA0est}) for fast purely magnetic waves. Note that since the density contrast satisfies that $0<\xi<1$ we always have that $u_R<c_{s01}<u_I$, and the same applies to $v_{A01}$.

Another important variable that is computed, once we know $v_0$, is the density enhancement/dimming associated to the incoming, reflected and transmitted wave. From Eq.~(\ref{densnonlincs}) we have that for the incoming wave
\begin{equation}\label{densnonlini}
    \rho_I=\rho_{01}\left(1+ \frac{\gamma-1}{\gamma+1}\frac{(u_I-u_R)(1+\xi)}{u_I(1-\xi)+u_R (1+\xi)}\right)^{2/(\gamma-1)},
\end{equation}
while for the dimming due to the reflection
\begin{equation}\label{densnonlinr}
   \rho_R=\rho_{01}\left(1- \frac{\gamma-1}{\gamma+1}\frac{(u_I-u_R)(1-\xi)}{u_I(1-\xi)+u_R (1+\xi)}\right)^{2/(\gamma-1)} ,
\end{equation}
where we have used again the expression for the reflection coefficient. The enhancement of the transmitted wave is
\begin{equation}\label{densnonlint}
    \rho_T=\rho_{02}\left(1+ 2\frac{\gamma-1}{\gamma+1}\frac{u_I-u_R}{u_I(1-\xi)+u_R (1+\xi)}\right)^{2/(\gamma-1)}.
\end{equation}
\noindent The expressions for the density fluctuations in the case of the magnetic case are given by Eqs.~(\ref{densnonlini})-(\ref{densnonlint}) but making the substitution $\gamma=2$ (instead of using $5/3$).

Finally, we derive an expression for the phase velocity of the transmitted wave into the CH in terms of the phase velocities of the incoming and reflected waves,
\begin{equation}\label{utcalc}
    u_T=\frac{1}{2 \xi}\left(u_I(1+\xi)+u_R(1-\xi)\right).
\end{equation}
This expression is the same for the purely magnetic case and for the purely acoustic case and we have used the fact that in our model $v_{A02}=v_{A01}/\xi$ and $c_{s02}=c_{s01}/\xi$. If measurements of incoming, reflected and transmitted phase speeds can be provided we are able to calculate the density contrast by using Eq.~(\ref{utcalc}). If in addition density measurements of the quiet Sun can be obtained from observations we are even capable of estimating the density inside of the CH.

In the following we apply actually measured values of incoming and reflected phase speeds ($u_I$, $u_R$) as well as measured density contrasts inside and outside the CH ($\rho_{02}/\rho_{01}$) in order to calculate the CW parameters $v_0$, $v_{A01}$ or $c_{s01}$, $\rho_I$, $\rho_R$, $\rho_T$ and $u_T$ by using the previous equations.

\subsection{Event 1}

 \citet{kienreichetal2013} analyses three homologous wave events referred to as W1, W2 and W3 in Table~\ref{tab:caption1} and Table~\ref{tab:caption2}) with clear reflection effects due to interaction with the same CH. The incident angle of the wave with respect to the CH normal is $\approx 10^{ \circ}$ meaning that the wave propagates almost perpendicularly to the CH boundary which is in agreement with the theoretical assumption made in Sect.~\ref{theory}. Incoming phase speeds and errors for the three primary waves are derived with $u_I=[155\pm 17, 180\pm 18, 219\pm 15]$ km $\rm s^{-1}$ while for the corresponding reflected waves the phase speeds are found to be $u_R=[119 \pm 28, 164\pm 33, 198\pm 34]$ km $\rm s^{-1}$ \citep{kienreichetal2013}. Since the phase speeds are close to the typical sound speed for a 1 MK corona we test the interpretation in terms of purely acoustic waves. For the calculations we use a density contrast of $\xi=\sqrt{0.43}$ and an error of $\pm0.02$, which has been derived from the density ratios using 193 and 195 \AA \:  EUV image data in Event 1 and Event 2 (considering $\rho/\rho_0 \sim \sqrt{I/I_{0}}$; see \citet{Zhukov2011}). The corresponding errors are calculated using the standard error-propagation formula. For the phase velocities we use the values given by the observations in Event 1 and Event 2. The calculated values for the velocities using Eqs.~(\ref{v0est})-(\ref{cs0est}), the corresponding Mach numbers and the transmitted phase speeds are found in Table~\ref{tab:caption1}. The acoustic and Alfv\'enic Mach numbers ($M$) are defined as the ratio of the velocity amplitude of the incoming wave to the sound speed and the Alfv\'en speed, respectively. The values for the densities calculated using Eqs.~(\ref{densnonlini})-(\ref{densnonlint}) are shown in Table~\ref{tab:caption2}. The estimated errors are also included in the tables.
 
  \subsection{Event 2}
 
 \citet{olmedoetal2012} also reported coronal waves reflected at a CH, although not caused by a strictly perpendicular incoming wave as the shape of the CH is rather complex. However, we chose this event since it is one of the most well-known cases for CW-CH-interaction giving measurements also for the transmitted wave of which there is in general a lack in the literature. The phase speed of the incident wave in this event is around 720 $\pm$ 20 km $\rm s^{-1}$, while the reflected wave propagates at 280  $\pm$ 10 km $\rm s^{-1}$. In this case the phase speed is closer to typical Alfv\'en speed values rather than to the sound speed, for this reason we give the estimation based on the magnetic interpretation only. \citet{olmedoetal2012} also found variations in the speed of the reflected wave, which shows that there might be projection effects and variations of the local speed. Secondary waves are reported to be deflected into the higher corona, which could also lead to a smaller projected speed \citep[see][]{kienreichetal2013}. The obtained values for the CW parameters are found in Tables~\ref{tab:caption1} and \ref{tab:caption2} (see Event 2).
 
% \textbf{Due to the higher values of the measured incoming and reflected wave, nonlinear effects are much stronger for this case compared to Event 1. - mat: please make more clear if calculated or measured!} \textit{It is clear that nonlinear effects are much stronger than in Event 1 since the values of the Mach numbers and density fluctuations for the incoming and transmitted waves are significantly higher.}} 
 \vspace{-15px}
\begin{table}[ht]
\begin{tabular}[t]{ccccc}
 \hline
 \makecell{Event} & \makecell{{$v_0$} \\ {(km $\rm s^{-1}$)}} & \makecell{{$c_{s0},v_{A0}$} \\ {(km $\rm s^{-1}$)}} & \makecell{Mach \\ number} &\makecell{{$u_T$}\\ {(km $\rm s^{-1}$)}}  \\
 \hline
 1 (W1) & 22$\pm$20 & 126$\pm$23 &  0.18$\pm$0.19 & 227$\pm$24 \\
 1 (W2) & 10$\pm$23 & 167$\pm$28 &0.06$\pm$0.15&270$\pm$26\\
 1 (W3) &13$\pm$23  & 202$\pm$28 & 0.06$\pm$0.12&328$\pm$23\\
 2  & 243$\pm$13 & 356$\pm$10&0.68$\pm$0.04&982$\pm$34\\
  \hline
\end{tabular}
\vspace{4px}
\centering
\caption{Calculated values for velocity amplitude of the incoming wave, sound speed, Alfv\'en speed, Mach number ($M$) and phase speed of the transmitted wave for Events 1 and 2.}
\label{tab:caption1}
\end{table}%
\vspace{-35px}

%For the purely acoustic case  we get for the velocity amplitudes $v_0=[23 \pm 22, 10 \pm 24 , 13\pm 24]$ km $\rm s^{-1}$ (Eq.~(\ref{v0estacoust})) and for the equilibrium sound speeds $c_{s01}=[124\pm 25 , 166\pm 29 , 201\pm 29]$ {km $\rm s^{-1}$} (Eq.~(\ref{cs0est})). The Mach numbers are $M=[0.18\pm 0.21,0.06 \pm 0.15,0.07\pm 0.13]$. For the density fluctuations we obtain $\rho_I/\rho_{01}=[1.20\pm 0.23, 1.06\pm 0.16 , 1.07\pm 0.13]$ (Eq.~(\ref{densnonlini})), while for the reflections, $\rho_R/\rho_{01}=[0.97\pm 0.06, 0.99\pm 0.03, 0.99\pm 0.03]$ (Eq.~(\ref{densnonlinr})). For the density changes in the transmitted wave we have that $\rho_T/\rho_{02}=[1.23 \pm 0.27, 1.07 \pm 0.19, 1.08\pm 0.16]$ (Eq.~(\ref{densnonlint})).
%For the purely magnetic case, using Eq.~(\ref{v0est}) we get for the velocity amplitudes $v_0=[20 \pm 19, 9 \pm 21, 12 \pm 21]$ km $\rm s^{-1}$ and for the equilibrium Alfv\'en speeds $v_{A01}=[124 \pm 25, 166 \pm 29, 201 \pm 29]$ km $\rm s^{-1}$ (same as the sound speeds). The Alfv\'enic Mach numbers are $M=[0.16 \pm 0.18, 0.05 \pm 0.14,0.06 \pm 0.11]$. For the density fluctuations, we have $\rho_I/\rho_{01}=[1.17 \pm 0.20, 1.05 \pm 0.14, 1.06 \pm 0.12]$, $\rho_R/\rho_{01}=[0.97 \pm 0.05, 0.99 \pm 0.03, 0.99 \pm 0.02]$, and  $\rho_T/\rho_{02}=[1.20 \pm 0.23, 1.06 \pm 0.16, 1.07 \pm 0.14]$.

\begin{table}[ht]
\begin{tabular}[t]{cccc}
\hline
\makecell{Event}  & \makecell{$\rho_I/\rho_{01}$}& \makecell{$\rho_R/\rho_{01}$}& \makecell{$\rho_T/\rho_{02}$}\\
\hline
1 (W1) &  1.19$\pm$0.21&0.97$\pm$0.04 & 1.23$\pm$0.26\\
 1 (W2) &  1.06$\pm$0.15& 0.99$\pm$0.03 & 1.07$\pm$0.19\\
 1 (W3) & 1.07$\pm$0.13&0.99$\pm$0.03 & 1.08$\pm$0.16\\
 2 &1.80$\pm$0.06&0.86$\pm$0.01 & 1.99$\pm$0.07\\
\hline
\end{tabular}
\vspace{4px}
\centering
\caption{Calculated values for density enhancement/dimming associated to incoming, reflected and transmitted wave for Events 1 and 2.}
\label{tab:caption2}
\end{table}%
 \vspace{-10px}
 
 From statistical studies we know that the density ratio lies between $0.1$ and $0.6$ \citep[e.g.,][]{Saqri2020,heinemann2019}. If we assume $\sqrt{0.1}\leq\xi\leq\sqrt{0.6}$, we are able to calculate upper and lower limits for the different parameters by using Eqs.~(\ref{v0est})-(\ref{utcalc}) and the limits for the density contrast, e.g, for Wave 1 in Event 1 we obtain $18\leq v_0\leq23$, $123\leq v_{A0}\leq131$, $0.14\leq  M \leq0.19$, $212\leq u_T\leq451$, $1.14\leq\rho_I/\rho_{01}\leq1.20$, $0.93\leq\rho_R/\rho_{01}\leq0.97$ and $1.22\leq\rho_T/\rho_{02}\leq1.23$. Analogously, parameter limits and therefore the dependence on the density contrast $\xi$ can be obtained for the other waves in both events.

 %The obtained values are $v_0=250 \pm 67$ km $\rm s^{-1}$, $v_{A01}=344 \pm 99$ km $\rm s^{-1}$, and the Mach number is 0.73 $\pm$ 0.40. For the density fluctuations we obtain, $\rho_I/\rho_{01}=1.86 \pm 0.55$, $\rho_R/\rho_{01}=0.88 \pm 0.14$ and $\rho_T/\rho_{02}=2.03 \pm 0.35$. The nonlinear effects are much more stronger than in Event 1. 
 
\section{Discussion and Conclusions}

We present a new and reliable method to calculate coronal wave parameters by using analytical expressions derived from linear wave theory and augmented by simple nonlinear terms of fast-mode MHD waves. The results have been validated by performing numerical simulations of CW-CH-interaction and have been applied to two different observational cases. With this we clearly emphasize the powerful combination between theory, simulations and observations.

The main results are summarized as follows:
\begin{enumerate}

\item We have applied the theoretical estimations to observations by calculating coronal wave parameters (e.g. density amplitudes, transmitted phase speed) by using incoming/reflected phase speeds and density contrast from the observations (see Eq.~(\ref{densnonlini}) - Eq.~(\ref{utcalc})).

\item We have performed numerical simulations of CW-CH-interaction and compared the results to the analytically and from linear theory derived reflection and transmission coefficients (see Eq.~(\ref{sr}) and Eq.~(\ref{st})). The obtained values show good agreement for the linear as well as the weakly nonlinear case, validating the method proposed in this Letter (see Figure \ref{retefig2}).

\item Moreover, if measurements of incoming, reflected and transmitted phase speeds are provided, the analytical expressions derived in this work can be used to obtain information about the CH itself, such as the density inside the CH and the density contrast to the surrounding (see Eq.~(\ref{utcalc})). 

\item Using the derived expressions for the local sound and Alfv\'en speeds (see Eq.~(\ref{cs0est}) and Eq.~(\ref{vA0est})), we are able to calculate the Mach numbers associated to the waves (see Table \ref{tab:caption1}). The large errors for these values can be explained by the uncertainties in the observed phase velocities.

\item Assuming we know the density contrast of the CH and its surrounding we are also able to calculate the Alfv\'en speed and the Mach number inside the CH.

\item If we know the density of the region of the incoming wave we are able to calculate the value of the magnetic field using the inferred Alfv\'en speed. 

\end{enumerate}

We have to keep in mind that we have considered a simplified model of the actual situation in the observations. In particular, we have studied a front that is perpendicular to the interface, which is not necessarily true in a real situation. The effect of the incident angle of the front needs to be taken into account in future studies. However, we have shown that theoretical estimations which were mainly derived from linear theory are a useful tool to calculate important coronal wave parameters in a fast and straightforward way, allowing us to perform coronal seismology.

%{\em Seismology: The estimated density, $\rho_{01}$, might be a %known magnitude and we can infer the magnetic field value from $ %B  =  \sqrt{\mu \rho_{01}} v_{A01}$.
%In our model the magnetic field is the same in the two regions %and we have an estimation of the density in the hole, then we %know the Alfv\'en speed in the CH, $v_{A02}  =  B/\sqrt{\mu %\rho_{02}}$.}

\begin{acknowledgements} We thank the anonymous referee for careful consideration of this manuscript and helpful comments. I.P. and J.T. acknowledge the support from grant AYA2017-85465-P
(MINECO/AEI/FEDER, UE), to the Conselleria d'Innovaci\'o, Recerca i Turisme del
Govern Balear, and also to IAC$^3$. This work was supported by the Austrian Science Fund (FWF): I 3955-N27.
\end{acknowledgements}
 \vspace{-30px}
\bibliographystyle{aa}      % basic style, author-year citations
\bibliography{letter}   % name your BibTeX data base

\begin{thebibliography}{21}
\expandafter\ifx\csname natexlab\endcsname\relax\def\natexlab#1{#1}\fi

\bibitem[{{Afanasyev} \& {Zhukov}(2018)}]{afanasyev2018}
{Afanasyev}, A.~N. \& {Zhukov}, A.~N. 2018, \aap, 614, A139

\bibitem[{{Gopalswamy} {et~al.}(2009){Gopalswamy}, {Yashiro}, {Temmer},
  {Davila}, {Thompson}, {Jones}, {McAteer}, {Wuelser}, {Freeland}, \&
  {Howard}}]{gopal2009}
{Gopalswamy}, N., {Yashiro}, S., {Temmer}, M., {et~al.} 2009, \apjl, 691, L123

\bibitem[{{Heinemann} {et~al.}(2019){Heinemann}, {Temmer}, {Heinemann},
  {Dissauer}, {Samara}, {Jer{\v{c}}i{\'c}}, {Hofmeister}, \&
  {Veronig}}]{heinemann2019}
{Heinemann}, S.~G., {Temmer}, M., {Heinemann}, N., {et~al.} 2019, \solphys,
  294, 144

\bibitem[{{Kienreich} {et~al.}(2013){Kienreich}, {Muhr}, {Veronig},
  {Berghmans}, {De Groof}, {Temmer}, {Vr{\v{s}}nak}, \&
  {Seaton}}]{kienreichetal2013}
{Kienreich}, I.~W., {Muhr}, N., {Veronig}, A.~M., {et~al.} 2013, \solphys, 286,
  201

\bibitem[{{Kienreich} {et~al.}(2011){Kienreich}, {Veronig}, {Muhr}, {Temmer},
  {Vr{\v{s}}nak}, \& {Nitta}}]{Kienreich2011}
{Kienreich}, I.~W., {Veronig}, A.~M., {Muhr}, N., {et~al.} 2011, \apjl, 727,
  L43

\bibitem[{{Landau} \& {Lifshitz}(1987)}]{landaulif1987}
{Landau}, L.~D. \& {Lifshitz}, E.~M. 1987, {Fluid Mechanics}

\bibitem[{{Liu} {et~al.}(2019){Liu}, {Wang}, {Lee}, \& {Shen}}]{liu19}
{Liu}, R., {Wang}, Y., {Lee}, J., \& {Shen}, C. 2019, \apj, 870, 15

\bibitem[{{Long} {et~al.}(2008){Long}, {Gallagher}, {McAteer}, \&
  {Bloomfield}}]{Long2008}
{Long}, D.~M., {Gallagher}, P.~T., {McAteer}, R.~T.~J., \& {Bloomfield}, D.~S.
  2008, \apjl, 680, L81

\bibitem[{{Mann}(1995)}]{mann1995}
{Mann}, G. 1995, Journal of Plasma Physics, 53, 109

\bibitem[{{Mihalas} \& {Mihalas}(1984)}]{mihalas1984}
{Mihalas}, D. \& {Mihalas}, B.~W. 1984, {Foundations of radiation
  hydrodynamics}

\bibitem[{{Muhr} {et~al.}(2011){Muhr}, {Veronig}, {Kienreich}, {Temmer}, \&
  {Vr{\v{s}}nak}}]{Muhr2011}
{Muhr}, N., {Veronig}, A.~M., {Kienreich}, I.~W., {Temmer}, M., \&
  {Vr{\v{s}}nak}, B. 2011, \apj, 739, 89

\bibitem[{{Olmedo} {et~al.}(2012){Olmedo}, {Vourlidas}, {Zhang}, \&
  {Cheng}}]{olmedoetal2012}
{Olmedo}, O., {Vourlidas}, A., {Zhang}, J., \& {Cheng}, X. 2012, \apj, 756, 143

\bibitem[{{Piantschitsch} {et~al.}(2018{\natexlab{a}}){Piantschitsch},
  {Vr{\v{s}}nak}, {Hanslmeier}, {Lemmerer}, {Veronig}, {Hernandez-Perez}, \&
  {{\v{C}}alogovi{\'c}}}]{Piantschitsch2018a}
{Piantschitsch}, I., {Vr{\v{s}}nak}, B., {Hanslmeier}, A., {et~al.}
  2018{\natexlab{a}}, \apj, 857, 130

\bibitem[{{Piantschitsch} {et~al.}(2018{\natexlab{b}}){Piantschitsch},
  {Vr{\v{s}}nak}, {Hanslmeier}, {Lemmerer}, {Veronig}, {Hernandez-Perez}, \&
  {{\v{C}}alogovi{\'c}}}]{Piantschitsch2018b}
{Piantschitsch}, I., {Vr{\v{s}}nak}, B., {Hanslmeier}, A., {et~al.}
  2018{\natexlab{b}}, \apj, 860, 24

\bibitem[{{Piantschitsch} {et~al.}(2017){Piantschitsch}, {Vr{\v{s}}nak},
  {Hanslmeier}, {Lemmerer}, {Veronig}, {Hernandez-Perez},
  {{\v{C}}alogovi{\'c}}, \& {{\v{Z}}ic}}]{Piantschitsch2017}
{Piantschitsch}, I., {Vr{\v{s}}nak}, B., {Hanslmeier}, A., {et~al.} 2017, \apj,
  850, 88

\bibitem[{{Saqri} {et~al.}(2020){Saqri}, {Veronig}, {Heinemann}, {Hofmeister},
  {Temmer}, {Dissauer}, \& {Su}}]{Saqri2020}
{Saqri}, J., {Veronig}, A.~M., {Heinemann}, S.~G., {et~al.} 2020, \solphys,
  295, 6

\bibitem[{{Veronig} {et~al.}(2006){Veronig}, {Temmer}, {Vr{\v{s}}nak}, \&
  {Thalmann}}]{Veronig2006}
{Veronig}, A.~M., {Temmer}, M., {Vr{\v{s}}nak}, B., \& {Thalmann}, J.~K. 2006,
  \apj, 647, 1466

\bibitem[{{Vr{\v{s}}nak} \& {Luli{\'c}}(2000)}]{vrsnaklulic2000}
{Vr{\v{s}}nak}, B. \& {Luli{\'c}}, S. 2000, \solphys, 196, 157

\bibitem[{{Walker}(2004)}]{walker2004}
{Walker}, A. 2004, {Magnetohydrodynamic Waves in Geospace}, Magnetohydrodynamic
  Waves in Geospace. Series: Series in Plasma Physics

\bibitem[{{Warmuth} {et~al.}(2004){Warmuth}, {Vr{\v{s}}nak}, {Magdaleni{\'c}},
  {Hanslmeier}, \& {Otruba}}]{Warmuth2004}
{Warmuth}, A., {Vr{\v{s}}nak}, B., {Magdaleni{\'c}}, J., {Hanslmeier}, A., \&
  {Otruba}, W. 2004, \aap, 418, 1101

\bibitem[{{Zhukov}(2011)}]{Zhukov2011}
{Zhukov}, A.~N. 2011, Journal of Atmospheric and Solar-Terrestrial Physics, 73,
  1096

\end{thebibliography}

\end{document}